\documentstyle[prl,twocolumn,aps,epsf]{revtex}

\begin{document} \draft
\twocolumn [ \hsize\textwidth\columnwidth\hsize\csname
@twocolumnfalse\endcsname
\title{Lattice Twisting Operators and Vertex Operators in Sine-Gordon
Theory in One Dimension}
\author{Masaaki Nakamura$^1$ and Johannes Voit$^2$}
\address{$^1$Max-Planck-Institut f\"{u}r Physik komplexer Systeme,
N\"{o}thnitzer Stra{\ss}e 38, 01187 Dresden, Germany}
\address{$^2$Theoretische Physik 1, Universit\"{a}t Bayreuth, 95440
Bayreuth, Germany}
\date{January 23, 2001}\maketitle
\begin{abstract}
\widetext\leftskip=0.10753\textwidth \rightskip\leftskip
 In one dimension, the exponential position operators introduced in a
 theory of polarization are identified with the twisting operators
 appearing in the Lieb-Schultz-Mattis argument, and their finite-size
 expectation values $z_L$ measure the overlap between the unique ground
 state and an excited state.  Insulators are characterized by
 $z_{\infty}\neq 0$. We identify $z_L$ with ground-state expectation
 values of vertex operators in the sine-Gordon model. This allows an
 accurate detection of quantum phase transitions in the universality
 classes of the Gaussian model. We apply this theory to the half-filled
 extended Hubbard model and obtain agreement with the level-crossing
 approach.
\end{abstract}
\pacs{71.10.Hf,71.10.Pm,75.10.Jm,77.84.-s}
] \narrowtext


The metal-insulator transition is a fundamental phenomenon in strongly
correlated electron systems. An insulator is distinguished from a
conductor at zero temperature by its vanishing dc conductivity (Drude
weight).  Kohn argued that localization of the electric ground-state
wave function is the signature of an insulating state\cite{Kohn}.
Recently, Resta emphasized that insulators are characterized by their
polarizability, and that meaningful definitions are required for the
polarization and position operators in extended systems. To this end, he
discussed the ground-state expectation value of the exponential of
position operators $\hat{x}_j$ in a finite size
system\cite{Resta,Resta-S1999,Resta2000},
\begin{equation}
\label{zdef}
z_L = \left\langle\exp\left(\frac{2\pi{\rm i}}{L}\sum_j\hat{x}_j\right)
\right\rangle\;.
\end{equation}
He showed that the many-body expectation value of position and
polarization operators in periodic systems are related to
$\frac{L}{2\pi}{\rm Im}\ln z_L$.  Generalization and numerical
calculations of $z_L$ were also
made\cite{Resta-S1999,Aligia-O,Aligia-H-B-O}, but our understanding of
this quantity is still far from satisfactory.  Specifically, we need to
know its relation to other theoretical schemes, and to clarify the
nature of phase transitions detected by this quantity.

In this Letter, we discuss the quantity $z_L$ from a different point of
view. Limiting ourselves to one-dimensional (1D) cases, we give two
interpretations to $z_L$: One is based on the argument by Lieb, Schultz,
and Mattis (LSM), which has been applied to spin and electron systems to
investigate structure of the excited
states\cite{Lieb-S-M,Affleck-L,Affleck,Oshikawa-Y-A,Yamanaka-O-A}.  The
other is the sine-Gordon theory\cite{Voit,Gogolin-N-T}, which describes
phase transitions in 1D quantum systems in terms of the renormalization
group. We will show that $z_L$ measures the orthogonality between the
unique ground state and a specific excited state of the finite size
system, and that it also gives the ground-state expectation value of a
vertex operator in the bosonization theory. Moreover, we show that the
condition $z_L=0$ corresponds to transition points that belong to the
universality class of the Gaussian model, and demonstrate this notion in
numerical analysis for a lattice electron model.


First, we discuss the new quantity based on a LSM-type argument using
the 1D spinless fermion model (SFM)
\begin{equation}
 {\cal H}=-t\sum_{i=1}^{L}
  (c^{\dag}_{i} c_{i+1}+\mbox{H.c.})+V(\{n_{i}\}),
\label{eqn:SF}  
\end{equation}
where the number of lattice sites $L$ is even, $c_{i}$ is the fermion
annihilation operator, and $V(\{n_{i}\})$ denotes the interaction term
given by arbitrary functions of the number operator $n_{i}\equiv
c^{\dag}_{i} c_{i}$.  The Fermi point is given by $k_{\rm F}=\pi N/L$
where $N$ is number of fermions.  To make the ground state
nondegenerate, we choose periodic (antiperiodic) boundary conditions for
$N=$ odd ($N=$ even), based on the Perron-Frobenious theorem.  Applying
to the ground state $|\Psi_0\rangle$ with momentum $k=0$ the unitary
``twisting operator'' ${\cal U}\equiv\exp[(2\pi{\rm i}/L)\sum_{j=1}^L j
n_j]$ $q$ times, generates a set of low-lying excited states
$|\Psi_q\rangle={\cal U}^q|\Psi_0\rangle$.  ${\cal U}$ is the lattice
version of the operator appearing in Eq.\ (\ref{zdef}). The translation
operator ${\cal T} (c_j\rightarrow c_{j+1})$ has eigenvalues ${\rm
e}^{{\rm i}k}$.  Due to the relation ${\cal U}^{-q}{\cal T}{\cal
U}^q={\cal T}{\rm e}^{{\rm i}2qk_{\rm F}}$, the operator ${\cal U}^q$
turns out to move $q$ fermions from the left Fermi point ($-k_{\rm F}$)
to the right one ($+k_{\rm F}$).  The excitation energy for
$|\Psi_q\rangle$ is evaluated as
\begin{eqnarray}
 \lefteqn{\Delta E=
  \langle\Psi_0|({\cal U}^{-q}{\cal H}{\cal U}^q-{\cal H})|\Psi_0\rangle}\\
 &=&-t[\cos(2q\pi/L)-1]\sum_{j=1}^L
    \langle\Psi_0|(c^{\dag}_{i} c_{i+1}+\mbox{H.c.})|\Psi_0\rangle.
    \nonumber
\end{eqnarray}
Thus, if the state $|\Psi_q\rangle$ is orthogonal to the ground state
$|\Psi_0\rangle$, there exists an excited state with energy of ${\cal
O}(L^{-1})$.

The orthogonality of these two states depends on the momentum of the
excited state $2qk_{\rm F}$. When $k_{\rm F}\neq\pi p/q$, where $p$ is
an integer and $p/q$ is an irreducible fraction, $|\Psi_0\rangle$ and
$|\Psi_q\rangle$ are characterized by different quantum numbers, so that
these two states are orthogonal.  On the other hand, when $k_{\rm F}=\pi
p/q$, the two states may not be orthogonal, so that this relation gives
a necessary condition for the system to have a
gap\cite{Oshikawa-Y-A,Yamanaka-O-A}.  The overlap of the two states in a
system of size $L$ is given by
\begin{equation}
   z_L^{(q)}
    \equiv\langle\Psi_0|\Psi_q\rangle=\langle\Psi_0|{\cal U}^q|\Psi_0\rangle.
    \label{eqn:def_z}
\end{equation}
$z_L$ is real due to parity invariance ($c_j\rightarrow c_{L+1-j}$). If
the system is gapped (gapless), $z_{\infty}\neq 0$ ($z_{\infty}=0$) is
expected. However, $z_L$ remains finite in finite-size systems even in
gapless cases. The expectation value of the twisting operator is nothing
but the quantity underlying Resta's definitions of electronic
localization and polarization\cite{Resta}, and its
generalizations\cite{Aligia-O}.


Next, we consider $z_L$ in a different point of view.  The low-energy
excitations of the SFM (\ref{eqn:SF}) with $k_{\rm F}=\pi p/q$ are
described by the sine-Gordon model\cite{Voit,Gogolin-N-T}
\begin{eqnarray}
   {\cal H}_{\rm SG}&=&\frac{v}{2\pi}\int_0^L {\rm d}x
  \left[K(\partial_x \theta)^2
       +K^{-1}(\partial_x \phi)^2\right]\nonumber\\
 &+&\frac{2 g}{(2\pi\alpha)^2}
  \int_0^L {\rm d}x \cos[q\sqrt{2}\phi(x)],
  \label{eqn:SG_model}
\end{eqnarray}
with a relation $[\phi(x),\theta(x')]=-({\rm i}\pi/2){\rm sign}(x-x')$.
This effective model consists of the Gaussian model and the nonlinear
term. Generally, the velocity $v$, the Gaussian coupling $K$, and the
Umklapp scattering amplitude $g$ are determined phenomenologically.  The
renormalization group (RG) equations for the nonlinear term are derived
under a change of the cutoff $\alpha\rightarrow {\rm e}^{{\rm
d}l}\alpha$\cite{Kosterlitz},
\begin{equation}
 \frac{{\rm d}y_{0}(l)}{{\rm d}l}=-y_1^{\ 2}(l),\ \ \
 \frac{{\rm d}y_1(l)}{{\rm d}l}=-y_{0}(l) y_1(l),
\label{eqn:RGE}
\end{equation}
where $y_{0}=2(q^2K/4-1)$, $y_{1}=g/\pi v$, and $l=\ln L$.

\begin{figure}[h]
\noindent
 \begin{center}
 \epsfxsize=2.0in \leavevmode \epsfbox{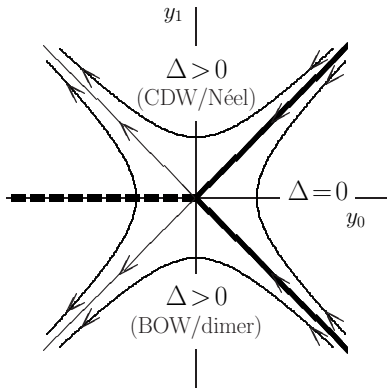}
 \end{center}  
\caption{RG flow diagram for the sine-Gordon model. Phase transitions
 described by this diagram are the WZNW type ($y_1=y_{0}=0$), the BKT
 type ($|y_1|=y_{0}>0$), and the Gaussian type ($y_1=0,y_{0}<0$).  The
 condition $z_L=0$ corresponds to the case $y_1=0$ which includes the
 WZNW-type and the Gaussian-type transitions.}  \label{fig:RGflow}
\end{figure}

The RG flow of this equation shown in Fig.~\ref{fig:RGflow} describes
the following three type of transitions: As is well known, the SFM can
be mapped on spin chains by a Jordan-Wigner transformation. When the
system has SU(2) symmetry, Eq.~(\ref{eqn:SG_model}) belongs to
universality class of the SU(2)$_1$ symmetric Wess-Zumino-Novikov-Witten
(WZNW) model\cite{Gogolin-N-T}, and the parameters are fixed as
$y_1=y_0$ ($q=2$). Then, when the initial value of $y_1$ is positive
$y_1(0)>0$, the nonlinear term is marginally irrelevant
[$y_1(\infty)=0$], and the system is gapless.  On the other hand, for
$y_1(0)<0$, the nonlinear term is relevant [$y_1(\infty)=-\infty$], a
gap opens, and the phase field is locked in the potential minimum as
$\sqrt{8}\langle\phi\rangle=0$ (mod $2\pi$).  We call the transition at
$y_1(0)=0$ ``WZNW type''.  In the U(1) symmetric case where the ratio of
$y_0$ and $y_1$ is no longer fixed, a Berezinskii-Kosterlitz-Thouless
(BKT) -type transition takes place at $|y_1(0)|=y_{0}(0)>0$.  For the
gapped states $|y_1(0)|>y_{0}(0)>0$, the parameters are renormalized as
$y_1(\infty)=\pm\infty$, and the phase field is locked as
$\sqrt{8}\langle\phi\rangle=\pi,0$ (mod $2\pi$).  In the SFM at
half-filling (spin chains with zero magnetic field), these two gapped
states correspond to the charge-density-wave (N\'{e}el) and the
bond-order-wave (dimer) states, respectively.  On the unstable Gaussian
fixed line [$y_1(0)=0$ with $y_{0}(0)<0$], a ``Gaussian transition''
takes place which is a second-order transition between the two gapped
states, and the system is gapless on the transition point.  These three
transitions can be identified by observing appropriate level crossings
in excitation spectra in finite-size systems\cite{Okamoto-N,Nakamura}.

Now, let us interpret $z_L$ in terms of the sine-Gordon theory.  In the
Tomonaga-Luttinger liquid theory which describes the Gaussian fixed
point ($y_1=0$), the current excitation created by the operator ${\cal
U}$ corresponds to the vertex operator $\exp({\rm
i}\sqrt{2}\phi)$\cite{Voit,Gogolin-N-T,Haldane}.  Besides, when $k_{\rm
F}=\pi p/q$, the phase field changes as $\phi\rightarrow-\phi$ under a
parity transformation, so that $z_L$ given by Eq.~(\ref{eqn:def_z}) is
related to the ground-state expectation values of the nonlinear term as
\begin{equation}
 z_L^{(q)}=\langle\cos(q\sqrt{2}\phi)\rangle.
  \label{eqn:relation}
\end{equation}
Since the sign of
the nonlinear term in Eq.~(\ref{eqn:SG_model}) changes at $y_1(0)=0$ in
the RG flow diagram, the WZNW-type and the Gaussian-type transition
points can be detected by observing $z_L=0$.  In the infinite-size
limit, we expect $z_{\infty}=\pm 1$ ($z_{\infty}=0$) for the gapped
(gapless) states.


To demonstrate the above argument, we consider the 1D extended Hubbard
model (EHM)
\begin{eqnarray}
  {\cal H}_{\rm EHM}&=&-t\sum_{i=1}^L\sum_{s=\uparrow,\downarrow}
 (c^{\dag}_{is} c_{i+1,s}+\mbox{H.c.})\\
 &+&U\sum_i n_{i\uparrow}n_{i\downarrow}
 +V\sum_i n_{i}n_{i+1},\nonumber
\label{eqn:tUV}
\end{eqnarray}
at half-filling and zero magnetic field ($k_{\rm F}=\pi/2$). Here $L$ is
even, and $c_{is}$ is the electron annihilation operator for spin
$s=\uparrow,\downarrow$. The number operators are defined by
$n_{is}\equiv c^{\dag}_{is} c_{is}$ and $n_i\equiv
n_{i\uparrow}+n_{i\downarrow}$.  The effective Hamiltonian density of
this system is given by sine-Gordon models for the charge ($\nu=\rho$)
and the spin ($\nu=\sigma$) sectors and a charge-spin coupling term
as\cite{Voit92}
\begin{eqnarray}
   \tilde{\cal H}&=&\sum_{\nu=\rho,\sigma}
   \frac{v_{\nu}}{2\pi}
  \left[K_{\nu}(\partial_x \theta_{\nu})^2
       +K_{\nu}^{-1}(\partial_x \phi_{\nu})^2\right]\label{eqn:eff_Ham}\\
 &+&\frac{2 g_{1\perp}}{(2\pi\alpha)^2}
  \cos[\sqrt{8}\phi_{\sigma}]
 -\frac{2 g_{3\perp}}{(2\pi\alpha)^2}
  \cos[\sqrt{8}\phi_{\rho}]\nonumber\\
 &+&\frac{2 g_{3\parallel}}{(2\pi \alpha)^2}
 \cos[\sqrt{8}\phi_{\rho}]\cos[\sqrt{8}\phi_{\sigma}].\nonumber
\end{eqnarray}
The spin part of this effective model belongs to the universality class
of the SU(2)$_1$ WZNW model.  According to the level-crossing
approach\cite{Nakamura}, the Gaussian transition in the charge part
($g_{3\perp}=0$), and the WZNW-type spin-gap transition ($g_{1\perp}=0$)
take place independently near the $U=2V$ line with $U>0$.  Therefore,
spin-density-wave (SDW), bond-charge-density-wave (BCDW), and
charge-density-wave (CDW) phases appear that correspond to the locked
phase fields
$(\langle\phi_{\rho}\rangle,\langle\phi_{\sigma}\rangle)=(0,*)$,
$(0,0)$, and $(\pi/\sqrt{8},0)$, respectively, where $*$ denotes the
unlocked (gapless) case.

To apply our argument to this electron system, we introduce the twisting
operators for the charge and the spin sectors following
Ref.~\ref{Yamanaka-O-A}
\begin{equation}
 {\cal U}_{\rho}\equiv {\cal U}_{\uparrow}{\cal U}_{\downarrow},\ \ \
 {\cal U}_{\sigma}\equiv {\cal U}_{\uparrow}{\cal U}_{\downarrow}^{-1},
\end{equation}
where ${\cal U}_{s}\equiv\exp[(2\pi{\rm i}/L)\sum_{j=1}^L j
n_{js}]$. Since this unitary operator corresponds to the vertex operator
given by $\exp[{\rm i}\sqrt{2}(\phi_{\rho}+s\phi_{\sigma})]$, we obtain
the expectation values of the nonlinear terms of Eq.~(\ref{eqn:eff_Ham})
as follows:
\begin{mathletters}\label{eqn:zzz}
\begin{eqnarray}
 z_{\nu}&\equiv&
 \langle\Psi_0|{\cal U}_{\nu}|\Psi_0\rangle
 = \mp\langle \cos(\sqrt{8}\phi_{\nu}) \rangle,\,\,\,\nu=\rho,\sigma,\\
 z_{\rho\sigma}&\equiv&
 \langle\Psi_0|
 {\textstyle\frac{1}{2}}({\cal U}_{\uparrow}^2+{\cal U}_{\downarrow}^2)
 |\Psi_0\rangle
 \nonumber\\
 &&
 =-\langle \cos(\sqrt{8}\phi_{\rho})\cos(\sqrt{8}\phi_{\sigma})\rangle.
\end{eqnarray}
\end{mathletters}
Therefore, $(z_{\rho},z_{\sigma},z_{\rho\sigma})$ are expected as
$(-1,0,0)$, $(-1,1,-1)$, and $(1,1,1)$ for the SDW, BCDW, and CDW
regions, respectively, in the $L\rightarrow\infty$ limit.  We calculate
these quantities in finite-size rings ($L=8$--$16$)
numerically. Eigenvalues and eigenstates of the EHM are obtained by the
Lancz\"{o}s algorithm and the inverse iteration method, respectively.

\begin{figure}[h]
\noindent
 \begin{center}
\epsfxsize=2.8in \leavevmode \epsfbox{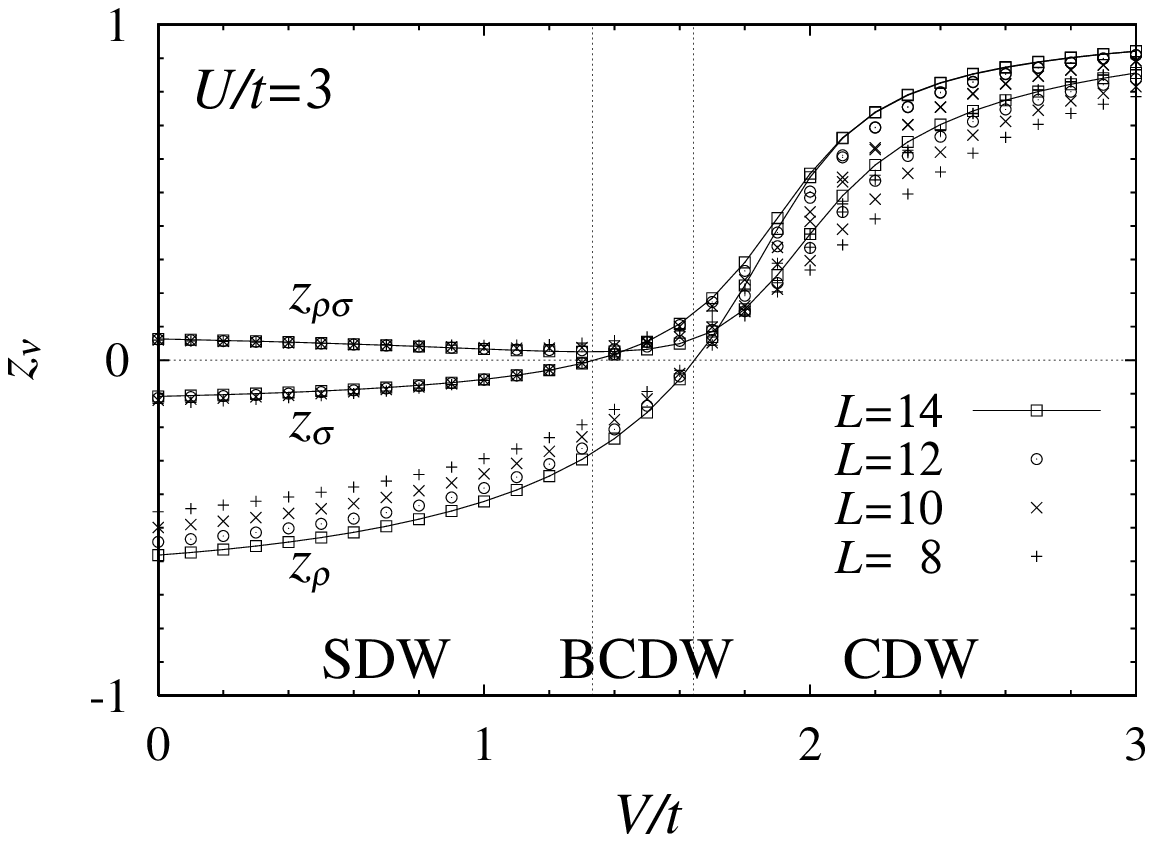}
 \end{center}
\caption{Behavior of $z_{\rho}$, $z_{\sigma}$, and $z_{\rho\sigma}$
 defined in Eq.~(\protect{\ref{eqn:zzz}}) at $U/t=3$ for $L=8,10,12$,
 and $14$ systems. In $L\rightarrow\infty$ limit,
 $(z_{\rho},z_{\sigma},z_{\rho\sigma})$ are expected as $(-1,0,0)$,
 $(-1,1,-1)$, and $(1,1,1)$ for the SDW, BCDW, and CDW regions.}
 \label{fig:zzz}
 \begin{center}
\epsfxsize=2.8in \leavevmode \epsfbox{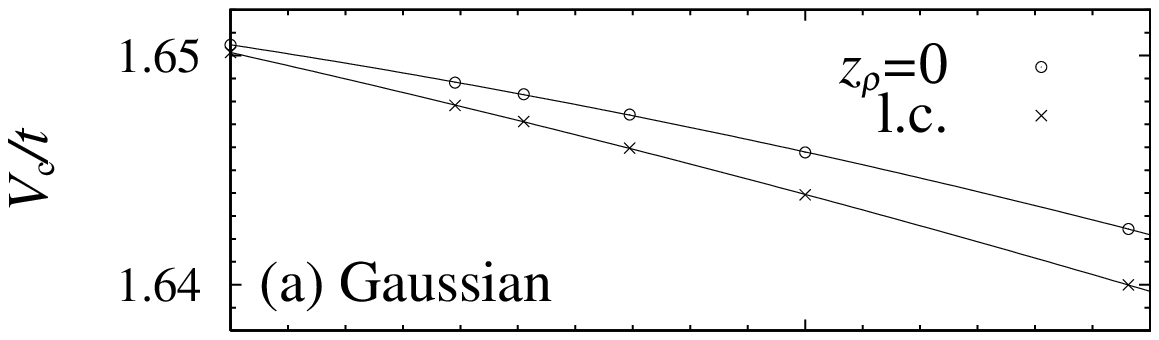}\\
\epsfxsize=2.8in \leavevmode \epsfbox{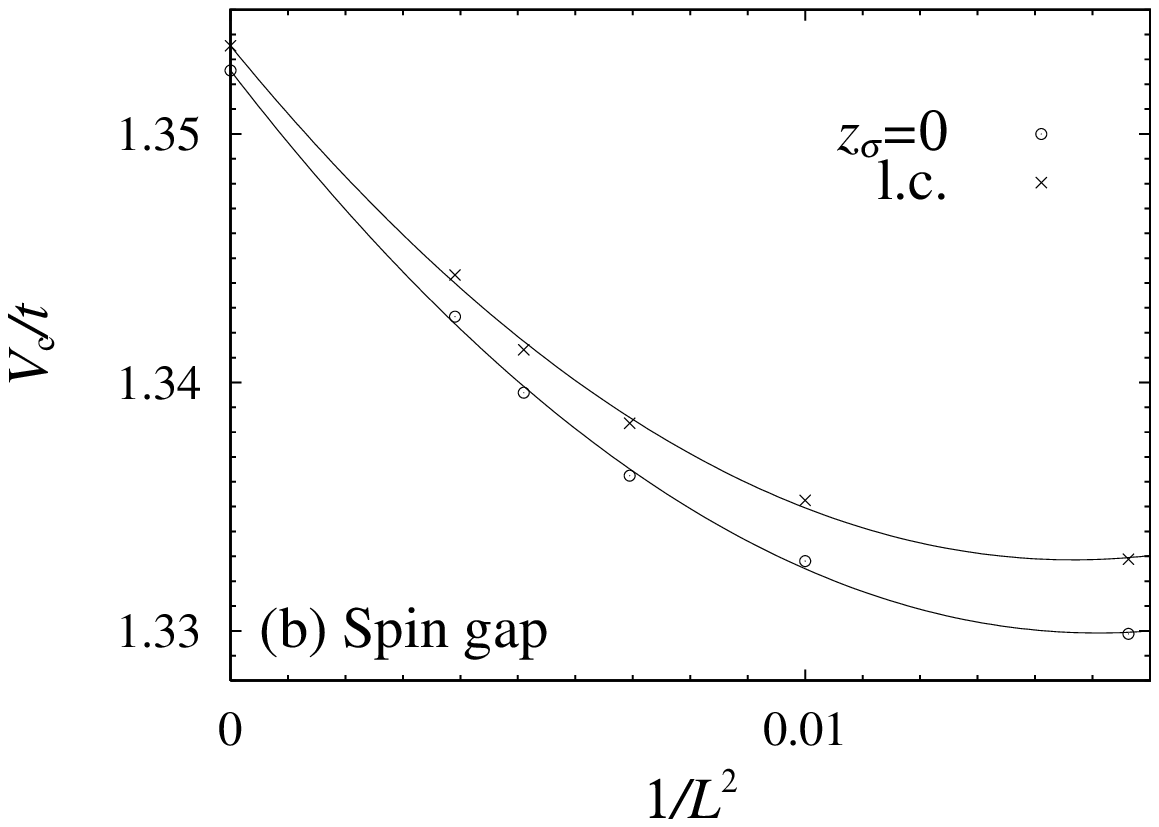}
 \end{center}
\caption{Size dependence of (a) the Gaussian transition point in the
 charge sector and (b) the spin-gap transition point at $U/t=3$,
 obtained by monitoring $z_{\nu}=0$ and the level
 crossings (l.c.).  The system sizes are $L=8,10,12,14$, and $16$. The
 extrapolation is done by $V_{\rm c}(L)=V_{\rm
 c}(\infty)+A/L^2+B/L^4$.}\label{fig:size_dep}
\end{figure}

In Fig.~\ref{fig:zzz}, we show the numerical results near the $U=2V$
line at $U/t=3$. For both charge and spin sectors, $z_{\rho}$ and
$z_{\sigma}$ change continuously, and change their signs at different
points: the Gaussian (BCDW-CDW) and the spin-gap (SDW-BCDW) transitions.
Although the convergence of $z_{\rho}$, $z_{\sigma}$, and
$z_{\rho\sigma}$ to their saturation values is slow except for the
transition points $z_{\nu}=0$, they are expected to take the predicted
values $0,\pm 1$, and change discontinuously at the transition points in
the $L\rightarrow\infty$ limit.

In Fig.~\ref{fig:size_dep}, we show the size dependence of the SDW-BCDW
and the BCDW-CDW boundaries at $U/t=3$ determined by the conditions
$z_{\nu}=0$. A benchmark for their accuracy is provided by the
level-crossing approach\cite{Nakamura} which is also shown.  These two
methods give quantitatively similar results, so that our interpretation
of the $z_L$ by the sine-Gordon theory is confirmed.  For extrapolating
the critical point to infinite system size, we use $V_{\rm c}(L)=V_{\rm
c}(\infty)+A/L^2+B/L^4$ given by conformal field theory, which is
justified when the non-linear sine-Gordon term is absent.  Numerical
results equivalent to ours were also obtained in Ref.~\ref{Torio-A-H-C}
by observing discontinuities of Berry phases explained in the following.
However, the physical interpretations are different. Besides, the
present calculation has great advantage in the accuracy and the
computational time.


We now discuss the quantity $z_L$ as a complex variable. $z_L$ can be
rewritten as a product of the discretized flux
state\cite{Aligia,Souza-W-M}
\begin{equation}
 z_L^{(q)}
  =\langle\Psi(0)|\Psi(2q\pi)\rangle
  =\prod_{j=0}^{M-1}\langle\Psi(\Phi_j)|\Psi(\Phi_{j+1})\rangle,
 \label{eqn:discrete}
\end{equation}
where $\Phi_j=2qj\pi/M$ with $M$ an integer.  Here $|\Psi(\Phi)\rangle$
is the ground state for the Hamiltonian ${\cal U}^{-1}(\Phi){\cal
H}{\cal U}(\Phi)$ with ${\cal U}(\Phi)\equiv\exp[({\rm
i}\Phi/L)\sum_{j=1}^L j n_j]$.  Now we consider the phase angle of $z_L$
in the complex plane $\gamma_{L}={\rm Im}\,\ln z_L$.  According to
Resta\cite{Resta}, this angle is regarded as the Berry phase
\cite{Resta2000,Aligia-H-B-O,Torio-A-H-C,Aligia,Souza-W-M,Zak1989,Ortiz-M,Resta-S1995}
\begin{equation}
 \gamma_{L}={\rm i}\int_0^{2q\pi} {\rm d}\Phi
  \langle \Psi(\Phi)|\partial_\Phi \Psi(\Phi)\rangle.
  \label{eqn:berry1}
\end{equation}
The angle $\gamma_{L}$ is $0$ or $\pi$ according to the signs of $z_L$,
so that it changes discontinuously at the Gaussian- or the WZNW-type
transition point [$y_1(0)=0$ with $y_0(0)\le0$].  However, it also shows
a discontinuity on the stable Gaussian fixed line [$y_1(0)=0$ with
$y_0(0)>0$], which does not correspond to any transitions (see
Fig.~\ref{fig:RGflow}).  In this case, $\gamma_{L}$ becomes meaningless
in the $L\rightarrow\infty$ limit.  For example, this situation appears
in the charge sector of the EHM, near the $U=2V$ line with
$U<0$\cite{Nakamura,Voit92}.  Extending $z_L$ into the complex plane may
be useful to distinguish phases with the same ${\rm Re\,}z_L$.  For
example, the SFM (\ref{eqn:SF}) at half-filling with
$V(\{n_{i}\})=\sum_i(Vn_{i}n_{i+1}+V'n_{i}n_{i+2})$ have a
charge-ordered state [$\langle n_{2k}\rangle=\langle
n_{2k+1}\rangle=[1+(-1)^k]/2$] for $V'\gg V>t$ that gives the same value
$z_{\infty}^{(2)}=-1$ as that of the bond-order-wave state [$\langle
n_{j}\rangle=1/2$].  We speculate that their Berry phases $\gamma_{L}$
are different, e.g. $\gamma_{L}=-\pi$ and $\gamma_{L}=\pi$. It would be
interesting to derive a simulation strategy (necessarily involving
parity breaking) to support such a conjecture.


In summary, we have given the following two interpretations to Resta's
expectation value of exponential position operators $z_L$: One is as
expectation values of a twisting operator which measures the
orthogonality between the unique ground state and an excited state
[Eq.~(\ref{eqn:def_z})].  The other is the ground-state expectation
value of the nonlinear term of the sine-Gordon model
[Eq.~(\ref{eqn:relation})].  From the latter point of view, it is shown
that insulating states characterized by $z_{\infty}=\pm 1$ correspond to
two different fixed points in the RG analysis, and that the condition
$z_L=0$ gives phase transition point which belongs to the Gaussian
universality class.  We have demonstrated this notion in the EHM, and
checked the validity of our argument comparing with the results of the
level-crossing approach.

Our work could open significant extensions: The present argument can be
applied to spin systems, with e.g. Haldane gaps and magnetization
plateaus.  The quantity $z_L$ may be applied to higher dimensional
cases, since the LSM argument has been extended to these cases by
considering a lattice wrapped on a
torus\cite{Lieb-S-M,Affleck,Oshikawa}.  Besides, the quantity $z_L$ can
be easily calculated numerically by the density matrix renormalization
group method or quantum Monte Carlo simulations, as it is based only on
ground-state properties.


One of the authors (M.N.) is grateful to M.~Arikawa for discussions.  He
also learned useful techniques for the exact diagonalization from
TITPACK Ver.~2 by H.~Nishimori.  The research of J.V.\ was supported by
Deutsche Forschungsgemeinschaft through grants no. VO436/6-2 (Heisenberg
fellowship) and VO436/7-1.

\end{document}